\documentstyle[twocolumn,psfig,rotating]{mn2e}
\def\lsim{~\rlap{$<$}{\lower 1.0ex\hbox{$\sim$}}}
\def\bsim{~\rlap{$>$}{\lower 1.0ex\hbox{$\sim$}}}

\def\kms{\ {\rm km\,s^{-1}}}
\def\hmpc{\ {\rm h^{-1}Mpc}}

\def\mdh{\ {\rm M_\odot/h}}
\def\hhmpc{\ {\rm h^{2}Mpc^{-3}}}

\def\la{\langle}
\def\ra{\rangle}

\def\etal{{\it et al.\ }}
\newcommand{\op}{Ly$\alpha$\ }
\newcommand{\hi}{\mbox{H{\scriptsize I}}}

\def\lya{\mbox{Ly$\alpha$\phantom{ }}}

\begin{document}
\title[The impact of galactic winds from LBGs on the Intergalactic Medium]
{The impact of galactic winds from LBGs on the Intergalactic Medium}
\author[Desjacques {\it et al.}]{Vincent Desjacques$^1$, 
Martin G. Haehnelt$^2$ and Adi Nusser$^1$ \\
$^1$Physics Department and Space Research Institute, Technion-Haifa\\
$^2$Institute of Astronomy, Madingley Rd, CB3-0HA, Cambridge, UK}
\maketitle

\begin{abstract}
An excess of sight-lines  close to Lyman-break galaxies (LBGs)  with
little or no absorption in QSO absorption spectra  has been 
reported and has been interpreted as the effect  of galactic winds 
on the Intergalactic Medium. We use here numerical  simulations to
investigate the flux probability function close to  plausible  sites
of LBGs. We show that the flux distribution near  our LBGs in the
simulation depends strongly on redshift, and is very sensitive to the
averaging procedure.  We show that a model without galactic winds and
a model with  a wind bubble size of $0.5 \hmpc$ (comoving) are
equally consistent with the new determination of the conditional flux
distribution by  Adelberger et al. (2005).  Models with the larger
bubble sizes ($\bsim 1\hmpc$) suggested by the previous observations
of Adelberger et al. (2003) based on a much smaller sample at higher
redshift are not consistent with the new data.  We, therefore, argue
that the volume  filling factor of galactic winds driven by LBGs may
be much smaller than previously thought and that most of the metals
responsible for the metal absorption associated with the low column
density \op forest  are unlikely to have  been ejected by LBGs.
\end{abstract}


\section {Introduction}
\label{introduction}

Feedback from supernovae is an important ingredient in scenarios of
galaxy formation. Observations of high-redshift galaxies  (Pettini
\etal 1998; 2001), nearby starbursts (Heckman \etal 1990)  and dwarf
galaxies (Heckman \etal 1995; 2001) suggest that  supernovae-driven
winds are ubiquitous in star-forming galaxies.  These could explain
among others the high temperature of the  high-redshift IGM (Cen \&
Bryan 2001), the faintness of the soft  X-ray background (e.g., Pen
1999) or the widespread detection of  metals in the spectra of quasars
(Cowie \etal 1995; Ellison \etal  1999; Schaye \etal 2000; Songaila
2001).

In the standard picture, galactic winds are driven by the energy
produced by supernova (SN) explosions of young massive stars.
However,  detailed modelling of a multi-phase ISM   remains
challenging (e.g. Ostriker \& McKee 1988; Efstathiou 2000)  and
predicting physical  properties  of galactic winds proves difficult.
Simulations incorporating simple phenomenological prescriptions  have
succeeded in producing SNe driven galactic winds  (Mac Low \& Ferrara
1999; Springel \& Hernquist 2002).  However, the results of  different
numerical studies often  do not agree (e.g., Croft \etal 2002; Theuns
\etal 2002), and the impact of winds on the IGM remains controversial.
In particular, the volume  fraction of the IGM affected by galactic
winds, the typical mass of the galaxies responsible for the enrichment
with the metals associated with the low column density \op forest,
and the redshift at which the enrichment occurred  remain a matter of
debate  (Gnedin 1998;  Cen \& Ostriker 1999; Aguirre \etal 2001a,
2001b;  Thacker, Scannapieco \& Davis 2002;  Bertone, Stoehr \& White
2004;  Haehnelt \& Pieri 2004; Aguirre et al. 2005; Pieri, Schaye \&
Aguirre 2005; Porciani \& Madau 2005; Rauch et al. 2005).

The Lyman-break technique (e.g. Steidel \etal 1996)  has enabled the
identification of large numbers  of  star-forming high-redshift
galaxies.  The large velocity shifts ($100-1000\kms$) between stellar
and  interstellar lines in the spectra of LBGs are strong evidence
that these galaxies drive powerful galactic winds  (Pettini \etal
1998, 2001; Heckman 2000). However, these spectroscopic measurements
do not tell us much about the location and volume filling factor of
the outflowing gas.  Measurements of the transmitted flux in QSO
absorption spectra with lines-of-sight  close to galaxies can provide
precious information  on the distribution  of ionized hydrogen in the
surroundings of these galaxies. Some of this gas has  presumably been
shock-heated by outflows.  In a seminal study,   Adelberger \etal
(2003, hereafter A03) measured the  flux of the \lya forest in the
absorption spectra of QSOs  as function  of the distance to the
nearest LBG.  At intermediate distances of 5-1 $\hmpc$ (comoving) they
found a decrease of the mean flux level as expected if LBGs are
located in overdense regions.  The few  (three) galaxies in their
sample lying closer than $0.5\hmpc$ to the line-of sight  showed,
however, a nearly complete lack of \op absorption  at the redshift of
the galaxies. These results  sparked off a series of  papers that
attempted to explain this by invoking feedback from ionizing radiation
(e.g. Maselli \etal 2004) or from galactic winds (Croft \etal 2002;
Bruscoli \etal 2003; Kollmeier \etal 2003a, 2003b; Desjacques \etal
2004; Kollmeier \etal 2005).   All these studies  had severe
difficulties to reproduce the observed large decrease of the
absorption at separation  $s\lsim 0.5\hmpc$ from the galaxies. Rather
large bubble sizes ($1\hmpc$ or larger) and extremely  efficient
feedback due to SN  (Croft \etal 2002)  and/or high star formation
rate  (Maselli \etal 2004) were required to explain the observations.
It was also pointed out that thermal motions and coherent peculiar
velocities strongly affect the galaxy-\op absorption correlation at
small separations (Kollmeier \etal 2003a).   Desjacques \etal 2004
showed  that these make it very difficult to explain the  rather
sudden increase of the mean flux level at a distance of $0.5\hmpc$.
 
Adelberger et al (2005,A05) have presented new results for the
transverse proximity effect of LBGs in QSO absorption spectra for a
larger sample at somewhat lower redshift than A03. In the new sample,
there are   24 galaxies at line-of-sight separations smaller than
$1\hmpc$. The sample is large enough to investigate  the conditional
probability distribution of the flux at this distance. The picture is
quite different to that of A03 where three out of three galaxies
showed very little absorption. The absorption close to these 24
galaxies covers the full range of flux levels but the PDF still shows
a peak at low absorption.   A05 have compared their measurements to
numerical simulations of Kollmeier et al.. They found good agreement
but an excess of objects  with little absorption. They attributed this
to the lack of galactic winds that were not modelled in the
simulations of Kollmeier et al.   We use here our own  simulations
with galactic winds similar to those presented in Desjaques et
al. (2004, D04) to investigate the implications of the new
observations  for the size of  the wind bubbles in our model.

\section{The model}
\label{sim}

\subsection{The \op forest and the galaxy distribution}

We  model the flux distribution of QSO absorption spectra  as a
function of the separation from the nearest LBG  as described in D04.
The modelling is based on a $\Lambda$CDM simulation whose  properties
are discussed in detail in Stoehr \etal (2002). Note  that we only
consider the particle distribution inside a box of  size $L=30\hmpc$
that belongs to a spherical region of high  resolution.  The particle
mass in this high-resolution region is $m=1.66\times  10^8\mdh$.  We
average the statistics of the simulated flux from  three snapshots at
redshift   $z=3$, 2.5 and 2, to facilitate a comparison with the
observations,  whose median redshift is $z=2.3$.

We associate LBGs with dark matter (DM) haloes identified with a
Friends-of-Friends group finding algorithm, assuming that  a DM halo
contains not more than  one LBG (e.g. Adelberger \etal 1998). Given
that LBGs have a wide range of stellar masses, and that the relation
between  LBGs and DM haloes is still somewhat uncertain, we have
investigated two  models as in D04.  In the ``starburst'' model, we
assume that the observed LBGs  reside in low mass haloes
(e.g. Lowenthal \etal 1997; Trager \etal  1997; Somerville \etal
2001).  We randomly select DM haloes such that  their number density
in the simulation is similar to the comoving  observed number density
$n_{_{\rm LBG}}$ of high-redshift LBGs.  We take $n_{_{\rm LBG}}$  to
be constant in the redshift range  $2\leq z\leq 3$, $n_{_{\rm
LBG}}=0.004\hhmpc$ (Giavalisco \&  Dickinson 2001).  In the
``massive-halo'' model, we assume that the high-redshift LBGs  are
the progenitors of the present-day massive and luminous galaxies
(e.g. Steidel \etal 1996;  Adelberger \etal 1998; Bagla 1998;
Haehnelt,  Natarajan \& Rees 1998). We pick a minimum  mass $M_{\rm
min}=5\times 10^{11}\mdh$ such that the number density of  haloes
$M\geq M_{\rm min}$ in the simulation is similar to the observed  LBG
density, $n_{_{\rm LBG}}$.

Synthetic spectra are extracted from the DM simulation in the usual
way. The spectra  are normalized to reproduce the mean transmitted
flux of  observed high-resolution spectra, $\la F\ra=0.68$, 0.79 and
0.85 at $z=3$, 2.5 and 2 respectively ({\it e.g.} McDonald \etal 2000).

\subsection{The effect of wind bubbles} 

Our simple model for  the effect of wind bubbles assumes that  shocks
produce  long-lived {\it fully ionised} spherical  bubbles around
galaxies (Croft \etal 2002; Kollmeier \etal 2003a,  2003b; Weinberg
\etal 2003; D04). We take the neutral hydrogen \hi\ density  to be
zero in a spherical region of radius $R$ centered on the  DM haloes
which we have identified  as hosts of LBGs. In the simulation, most of
the haloes lie along  filaments and the gas distribution  is often
anisotropic (e.g. Croft \etal 2002, Springel \& Hernquist
2003). Theuns \etal (2002) have also shown that winds propagate more
easily into the low density IGM than in the filaments.  We therefore
caution that our
assumption  of a spherical wind bubble will  only be  a reasonable
approximation for strong isotropic winds.  Furthermore wind bubbles do
not necessarily live for a Hubble  time.

\begin{figure}
\resizebox{0.43\textwidth}{!}{\includegraphics{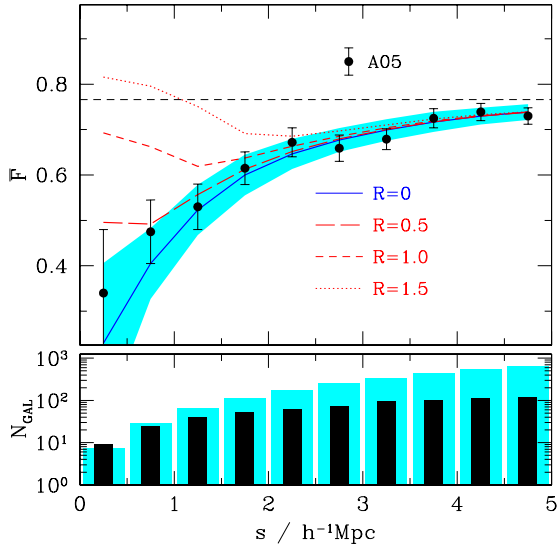}}
\caption{The mean flux ${\bar F}$  as a function of the distance $s$
to the nearest halo in the starburst model for different radii of the
wind bubble, 0.5, 1 and 1.5$\hmpc$  (top panel).  The solid curve
shows ${\bar F}$ for a model without galactic winds ($R=0$).  The
horizontal dashed line shows the mean transmissivity  $\la f\ra$=0.765
of the observed sample. The shaded area is our cosmic  variance
estimate for the model without galactic winds  calculated
from  the cumulative distribution $N_{\rm GAL}$ shown as the light
(blue) histogram at the bottom.  The filled symbols (top) and
the dark bars (bottom) show   ${\bar F}$ and $N_{\rm GAL}$ of the
observed sample of A05.}
\label{fig:fbar}
\end{figure}

\section{Comparison with observations}
\label{compare}

\subsection{Cosmic variance and galaxy redshift uncertainty}
\label{error}

The new sample of A05 is significantly larger than that of A03. At
separation $s\leq 1\hmpc$, there are 24 galaxies with good redshift
determination. Even though the new sample is larger than the A03
sample, it is still crucial to assess the statistical errors on the
derived flux probability distribution  resulting from the finite
number of galaxies and QSO spectra.   We focus on the conditional PDF
$P({\bar D},s<1)$ of the flux decrement ${\bar D}=1-{\bar F}$ at a
separation $s\leq 1\hmpc$, and the conditional transmitted flux ${\bar
F} (s)$.  We estimate the statistical errors  using Monte-Carlo
realizations of the flux distribution.  We found that a sample of
about 80 LOS through our simulation  box gives a cumulative ``galaxy''
function $N_{\rm GAL}(<s)$ close to that of the observed sample at
separation $s \leq 1\hmpc$ (see Fig.~\ref{fig:fbar}).  The simulated
sample has 28 pairs with separation  $s\leq 1\hmpc$, while the
observed sample has 24 pairs on that scale.  The discrepancy at large
scales is due to the  (small) size of the observed fields.  The number
of haloes with mass  $M>10^{11}\mdh$ in the   simulation is about
1000. We should thus not have oversampled  the halo catalogue  too
much. To account for redshift errors,  we add a Gaussian error with
variance $\Delta v=60\kms$.

\subsection{The conditional transmitted flux}
\label{flux}

Fig.~\ref{fig:fbar} compares the conditional flux of our  starburst
model with  the measurements of A05, shown as filled symbols. The
assumed bubble radius is $R=0$ (solid), 0.5 (long dashed), 1 (short
dashed)  and 1.5 $\hmpc$ (dot), respectively.  ${\bar F}$ has been
computed from the full three-dimensional distribution  of the  flux
distribution in the simulation. Results obtained from the various
snapshots have been rescaled to the mean transmissivity $\la
f\ra=0.765$ of the observed  sample (shown as the horizontal dashed
line) before averaging. The shaded area shows the 1$\sigma$ error for
the $R=0$ model  computed from the cumulative halo distribution
plotted as the light (blue) histogram in the bottom panel. On scales
$s\lsim 1\hmpc$ our estimate of cosmic variance is consistent  with
that of A05, though somewhat larger by $\sim$20 per cent.
Fig.~\ref{fig:fbar} shows that the $R=0$ and $R=0.5$ models are
consistent with the data of A05, while the models with $R=1$ and 1.5
significantly overestimate ${\bar F}$ at separation $s\lsim 1\hmpc$.

\subsection{The conditional probability distribution}
\label{pdf}

We calculate $P({\bar D}|s<1)$ and its cosmic variance error using
Monte-Carlo realizations of 28 sight lines, where ${\bar D}$ is the
mean flux decrement of all pixels within $1\hmpc$ from a LBG. We
assume that there is one sight line per LBG. We normalize  $P({\bar
D}|s<1)$ such that the observed PDF gives the observed  number of
galaxies in each bin.

\begin{figure}
\resizebox{0.43\textwidth}{!}{\includegraphics{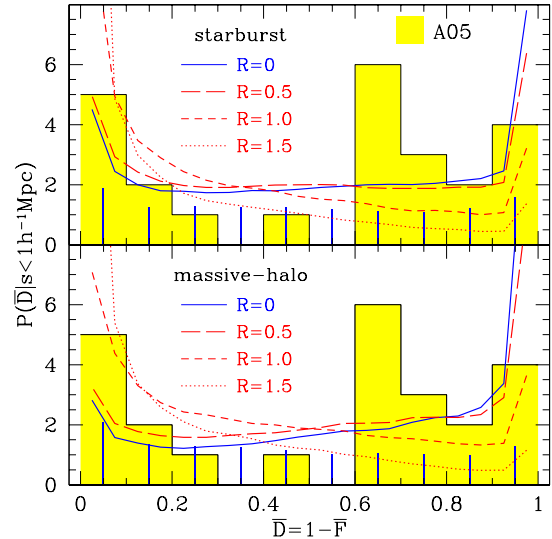}}
\caption{The differential probability distribution of the flux
decrement ${\bar D}$ at a separation $s<1\hmpc$ from a LBG, in the
starburst and massive-halo models (top and bottom respectively).
$P({\bar D}|s<1)$  is plotted for the models of Fig.~\ref{fig:fbar}.
The histogram shows the measurements of A05. The vertical bars
indicate the cosmic variance in bins of $\Delta{\bar D}=0.1$ for the
model $R=0$.}
\label{fig:pdf1}
\end{figure}

\subsubsection{The conditional PDF in the starburst model}
\label{starburst}

In Fig.~\ref{fig:pdf1} we compare the conditional PDF of the models
plotted in Fig.~\ref{fig:fbar} with the measurements of A05. The
narrow vertical bars indicate the 1$\sigma$ error (in bins of
$\Delta{\bar D}=0.1$ for the  model  with $R=0$). All models show a
broad distribution whith peaks at  small and large flux
decrements. The Kolmogorov-Smirnov (KS) probability that the observed
distribution is drawn from the same PDF as our simulated distribution
is highest for the model without galactic winds (0.315). It is still
reasonably high for the model with $R=0.5$ (0.130), but decreases
sharply for the models with larger bubble radii ($\lsim 10^{-3}$). The
rather large errors suggest that statistical fluctuations could  be
responsible for the lack of LBGs with flux decrements in the range
${\bar D}\sim 0.3-0.6$ as well as the peak at ${\bar D}=0.65$, which
is  roughly 3$\sigma$ above the mean. The existence of LBGs with
${\bar D}<0.2$ in the $R=0$ model may be surprising. One may have
expected  most galaxies to have ${\bar D}\lsim 1$ as they reside in
dense regions with significant amounts of neutral hydrogen.  Note,
however, that the   \hi\ density  decreases rapidly   away from
LBGs. In fact, we found that $\bsim 55$ per  cent of the LBGs with
${\bar D}<0.2$ have a impact parameter  $\geq 0.8\hmpc$. This fraction
is only $\sim 35$  per cent averaged over all flux decrements.
Lines-of-sight with weak absorption have preferentially a large impact
parameter in the $R=0$ model. Note that this trend appears not to be
present in the observations, where only 25  per cent of the LBGs with
${\bar D}\lsim 0.2$ have an impact parameter  $\geq 0.8\hmpc$ (Table 3
of A05).

\subsubsection{The conditional PDF in the massive-halo model}
\label{massive}

The sensitivity of $P({\bar D}|s<1)$ to the bubble radius $R$ suggests
that it may also  depend on the properties of the haloes hosting the
LBGs. The bottom panel of Fig.~\ref{fig:pdf1} demonstrates that this
is indeed the case. The conditional PDF is shown for our massive-halo
model with the same bubble radii as for the starburst model. Again,
all models have a broad distribution with peaks at small and large
flux decrements similar to that of the starburst model. However,  the
fraction of LBGs with flux decrement ${\bar D}<0.2$ is somewhat
smaller. This is most likely due to  the increase of the typical
infall velocities with increasing  halo mass (D04).  In the
massive-halo scenario, the KS probability is highest  for the model
with bubble radius $R=0.5\hmpc$ (0.151). The probability is only 0.014
in the   $R=0$ model, and $\ll 0.01$ for the models with $R\geq
1\hmpc$.  A significant fraction of   ionised bubbles with  $R\sim
0.5\hmpc$ is thus required in the massive-halo scenario to reproduce
the observed  distribution. Note, however,   that the rather large
errors make  the  interpretation of the KS probabilities problematic.
Our results are consistent with our previous study where we also found
that models  where  LBGs are mainly starbursts in small mass haloes
matches the observations with  smaller bubble radii than models where
massive haloes host the LBGs (D04).

\begin{figure}
\resizebox{0.43\textwidth}{!}{\includegraphics{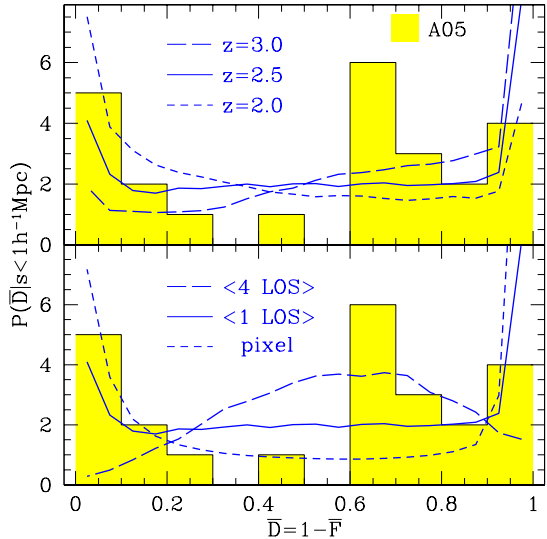}}
\caption{The differential probability distribution at separation
$s<1\hmpc$  from a LBG as a function of the flux decrement ${\bar D}$
in the starburst  model without galactic winds ($R=0$).  The top panel
shows the  redshift dependence, and the bottom panel demonstrates the
sensitivity to the averaging procedure (see text for  details). The
histogram is the data of A05.}
\label{fig:pdf2}
\end{figure}

\subsubsection{Sensitivity to  redshift and averaging procedure}
\label{sys}

The conditional probability of the models shown in Fig.~\ref{fig:pdf1}
is an average of three equally weighted snapshots at $z=3$, 2.5 and 2.
However,  $P({\bar D},s<1)$ evolves noticeably over that redshift
range.  This is shown in the top panel of Fig.~\ref{fig:pdf2}, where
$P({\bar D},s<1)$ is plotted as a function of redshift for the
starburst model with $R=0$. Measurements  at $z=2$ contribute most to
the low flux decrement tail of the PDF. The predicted mean number of
galaxies with ${\bar D}<0.2$ is $n_{<0.2}=2.6$, 5.0 and 8.6 in a
sample with 24 objects at $z=3$, 2.5 and 2, respectively. In the
massive-halo model,  these values drop to $n_{<0.2}=1.7$, 3.4 and 5.1,
respectively. The strong dependence of $P({\bar D},s<1)$ on redshift
reflects the evolution of the unconditional PDF of the \op flux in
that redshift interval.  The selection function of the A05 sample
peaks at $z=2.3$.  Among the 24 galaxies with impact parameter $\leq
1\hmpc$, only  5 have a redshift $z>2.5$.  We have also tried a
weighting which matches the observed  redshift distribution more
closely and got similar results.

The bottom panel of Fig.~\ref{fig:pdf2} demonstrates that the shape of
the conditional PDF is also strongly sensitive to the averaging
procedure  used to calculate ${\bar D}$. The solid curve is for our
default  assumption, and is computed as in A05. To obtain the long
dashed curve, we averaged over four lines-of-sight  per LBG.  The
short dashed curve  shows the PDF obtained if pixels in the spectrum
are considered individually. Averaging  over four lines-of-sight  per
LBG moves the flux  decrement in the tails of the distribution closer
to the mean.  When all pixels are considered separately the peaks at
small and large  flux decrement are enhanced. These findings are
consistent with the results of Kollmeier \etal (2003a).

\section{Summary and discussion}
\label{discussion}

We have revisited the proximity effect of LBGs on the  observed flux
distribution in QSO absorption spectra and have investigated the
implications of the new data of A05 for the bubble size and volume
filling factor of galactic winds of LBGs. We have used  mock spectra
calculated from \op forest simulations   with the galactic wind
prescription of D04 to model  the conditional probability distribution
of the flux as a function of the line-of-sight distance to the closest
LBG. As in D04 we have considered two scenarios for the LBGs, a
starburst model and a massive-halo model.  We have used Monte-Carlo
realizations of the flux distribution   to assess the statistical
uncertainties.    Our main results are as follows.

\begin{itemize}

\item{Our simulations give a broad probability distribution of the
flux  for lines-of-sight passing close to LBGs ($s\leq 1\hmpc$ ) with
a peak at small and large flux decrements consistent with that
observed by A05 within the expected errors for a sample of this size.}

\item{None of the models leads to the observed  lack of galaxies with
decrement $0.3\leq {\bar D}\leq 0.6$, or the observed  excess at
${\bar D}\sim 0.7$. However, error bars are large and statistical
fluctuations may be responsible for  these features.}

\item{For the starburst model the conditional PDF of the model with no
galactic winds  agrees best with the observations, but the model with
a bubble radius $R=0.5 \hmpc$ is also consistent.  For the massive
halo model the model with  $R=0.5 \hmpc$  agrees best with the
observation}.

\item{The conditional PDF of the flux depends strongly on redshift and
the averaging procedure applied.}

\end{itemize}

Our results appear  to disagree somewhat with the prediction  of the
SPH simulation without galactic winds shown in Fig. 15 of the A05
paper.  Note, however, that the latter was obtained from a single
output at $z=3$ rescaled to the mean redshift of the observed sample.
The strong dependence of $P({\bar D},s<1)$ on redshift could explain
the absence of the bimodality in the PDF of the  SPH simulation. Of
course, our simplified treatment of the IGM physics may also be the
culprit for the difference.

The good agreement of our starburst model without galactic winds  with
the measurements of A05 suggests that the volume fraction of the  IGM
affected by winds may be very small.  A model where LBGs drive
outflows extending substantially less than  $0.5\hmpc$ (comoving) can
still explain the peak  in the conditional PDF  at small flux
decrement, and reproduce the mean flux level as a function of the
distance to the closest LBG, irrespective  of the exact properties of
the DM haloes hosting LBGs.  Note that in such a model galactic winds
from LBGs, while still being responsible  for the observed strong CIV
absorption close to these galaxies (A05),  would nevertheless have
very little effect on global statistical properties  of  the \op
forest (e.g. Theuns \etal 2002; Desjacques \etal 2004;  McDonald \etal
2005). They would also not be responsible for the majority of the
weaker CIV absorption which has a much larger volume filling factor.

\section{Acknowledgments}

We acknowledge stimulating discussions with Juna Koll\-meier and Ehud
Behar. We thank Felix Stoehr for providing us with his numerical 
simulations.  This Research was supported by the Israel Science
Foundation (grant No. 090929-1) and the EC RTN network ``The Physics 
of the IGM''.

\end{document}